\documentclass[a4paper,11pt]{article}
\pdfoutput=1 

\usepackage{jinstpub} 

\usepackage{lineno}

\title{\boldmath CMS RPC Background - Studies and Measurements}

\author[d,1]{R. Hadjiiska,\note{Corresponding author.}}
\author[a]{A. Samalan,}
\author[a]{M. Tytgat,}
\author[a]{N. Zaganidis,}
\author[b]{G.A. Alves,}
\author[b]{F. Marujo,}
\author[c]{F. Torres Da Silva De Araujo,}
\author[c]{E.M. Da Costa,}
\author[c]{D. De Jesus Damiao,}
\author[c]{H. Nogima,}
% ...

\affiliation[a]{Ghent University, Dept. of Physics and Astronomy, Proeftuinstraat 86, B-9000 Ghent, Belgium}
\affiliation[b]{Centro Brasileiro Pesquisas Fisicas, R. Dr. Xavier Sigaud, 150 - Urca, Rio de Janeiro - RJ, 22290-180, Brazil}
\affiliation[c]{Dep. de Fisica Nuclear e Altas Energias, Instituto de Fisica, Universidade do Estado do Rio de Janeiro, Rua Sao Francisco Xavier, 524, BR - Rio de Janeiro 20559-900, RJ, Brazil}
\affiliation[d]{Bulgarian Academy of Sciences, Inst. for Nucl. Res. and Nucl. Energy, Tzarigradsko shaussee Boulevard 72, BG-1784 Sofia, Bulgaria}
% ...

\emailAdd{roumyana.mileva.hadjiiska@cern.ch}

\abstract{The expected radiation background in the CMS RPC system has been studied using the MC prediction with the CMS FLUKA simulation of the detector and the cavern. The MC geometry used in the analysis describes very accurately the present RPC system but still does not include the complete description of the RPC upgrade region with pseudorapidity $1.9 < \lvert \eta \rvert  < 2.4$. Present results will be updated with the final geometry description, once it is available. The radiation background has been studied in terms of expected particle rates, absorbed dose and fluence. Two High Luminosity LHC (HL-LHC) scenarios have been investigated - after collecting $3000$ and $4000$ fb$^{-1}$. Estimations with safety factor of 3 have been considered, as well.}

\keywords{Resistive-plate chambers, Radiation monitoring, Radiation calculations}

\collaboration[c]{on behalf of the CMS collaboration}

\proceeding{15$^{\text{th}}$ Workshop on Resistive Plate Chambers and related detectors\\
  10-14 Feb 2020\\
  Rome, Italy}

\begin{document}
\maketitle
\flushbottom

\section{Introduction}
\label{sec:intro}
\begin{figure}[htbp]
	\centering
	\includegraphics[width=.6\textwidth]{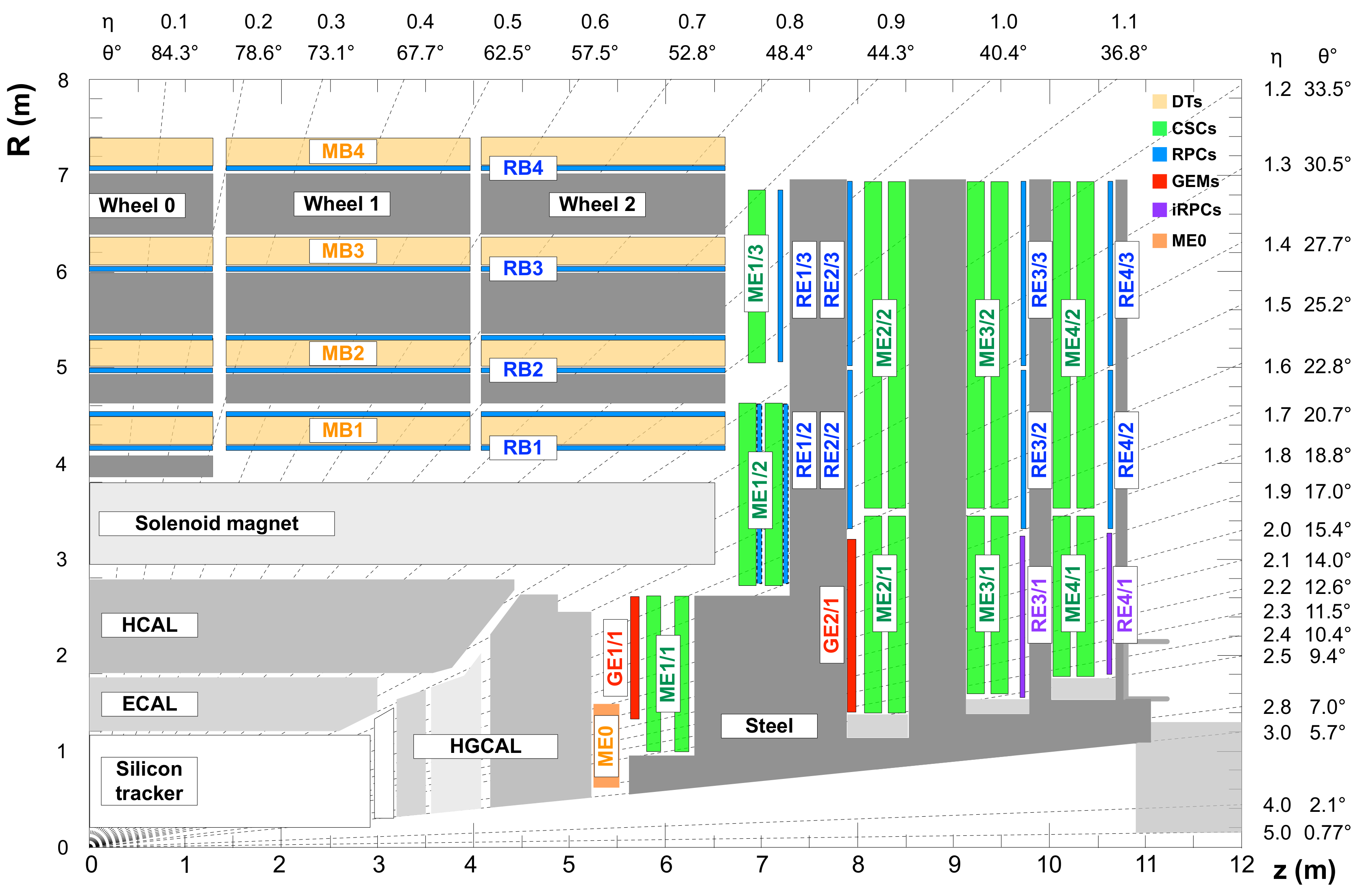}
	\qquad
	\caption{\label{fig:muonsys} A CMS quadrant plot with Phase-2 Muon Upgrade is shown on the figure. The RPC stations are labeled with RBX in the barrel and REX/Y in the endcap, where X depicts the station, and Y the ring enumeration. The iRPC chambers will be installed on RE3/1 and RE4/1. }
\end{figure}

A quadrant plot of the CMS (Compact Muon Solenoid) detector is shown in Fig. \ref{fig:muonsys}. CMS is composed of a barrel and two endcap regions. A detailed description of the CMS detector can be found in Ref. \cite{cms}. The CMS muon system consists of five separate wheels in the barrel and four disks for each of the two endcaps. Each wheel has four different stations (concentric layers) and each endcap disk is organized in three rings. Three different gaseous detector technologies are used in the present muon system - DT (Drift Tubes) in the barrel, CSC (Cathode Strip Chambers) in the endcap and RPC (Resistive Plate Chambers) both in barrel and endcap. One of the Phase-2 Muon Upgrade subjects is the installation of additional detector layers in the high pseudorapidity region: GEMs (Gas Electron Multipliers) on the inner ring of the first and second disk, ME0 station which will be composed of six layers of GEM detectors, and so-called iRPCs (improved RPC) on the inner ring of the third and fourth disk \cite{muontdr}. The study of the existing and new potential radiation background sources is important for the choice of detector technologies and electronics components. The CMS RPC system is subject to two main radiation sources: the first one from collisions and radiation leaks through HCAL (Hadron Calorimeter), affecting mainly the first stations of the external barrel wheels, and the second one due to the neutron-induced background from the cavern, which affects mainly the outermost endcap disks and the fourth barrel stations. The expected radiation background in the CMS RPC system has been studied using the MC prediction with the CMS FLUKA \cite{fluka1,fluka2} simulation of the detector and cavern. The radiation background has been studied in terms of expected particle rates, absorbed dose and fluence. Two High Luminosity LHC (HL-LHC) scenarios have been investigated - after collecting $3000$ and $4000$ fb$^{-1}$.\\
This work is focused on the radiation background on the RPCs and iRPCs of the third and fourth endcap disks. The present RPC system has chambers installed on the second and third rings of each disk. During the Phase-2 upgrade the iRPC chambers will be installed on the inner rings of the third and forth disks. In Fig. \ref{fig:muonsys} the third and fourth RPC endcap disks are indicated as RE3/Y and RE4/Y, where Y depicts the ring enumeration ($Y = 1,2,3$). RE3/1 iRPCs will be mounted directly on the endcap yoke. The FEBs (Frond-End Boards) will be mounted behind the RE3/1 chambers. RE4/1 iRPCs will be installed in a high pseudorapidity region over the existing CSC chambers (ME4/1 in the same figure). More details about the integration and installation of the iRPC can be found in Ref. \cite{voevodina}.

\section{Validation with experimental data}
\label{sec:realdata}
During the LHC Run-2 $(2015-2018)$, the experimental RPC rates have been studied as a function of the instantaneous luminosity. The rates are obtained from the Link-board counts \cite{silvia15} (data transfer boards from FEB to trigger system) of the present chambers. Extrapolation to a higher luminosity scenario can be done based on the already observed linear dependence between the RPC hit rates and the LHC instantaneous luminosity (see Ref. \cite{raul}). The maximum rate is  for the top RE4/2 chambers (installed at $80^{\circ} < \phi < 100^{\circ}$, where $\phi$ is the azimuthal angle) and it is expected to be  $\sim200$ Hz/cm$^{2}$.  To evaluate the performances and radiation hardness required to the detector we assume a safety factor of 3 (SF(3)) so that we get a maximum rate of about 600 Hz/cm${^2}$.

Experimentally derived data have been used also to validate MC simulation with FLUKA. Experimental rates measured at an instantaneous luminosity of $1.5\times10^{34}$ cm$^{-2}$s$^{-1}$ have been compared to the MC results with a dedicated CMS Run-2 FLUKA simulation. As it is reported in Ref. \cite{silvia20}, a good agreement between the experimental and MC results is observed for the barrel part (up to $|\eta| < 1$), where $\eta$ is the pseudo-rapidity. Some discrepancy is observed in the endcap ($1 < |\eta| < 2.4$), mainly in the overlap region between barrel and endcap ($1 < |\eta| < 1.2$), where MC predicts higher rates. It is caused by some differences in the geometry model with respect to the real detector. On average the discrepancy in the entire endcap region is $2$, but the choice of safety factor $3$ overcomes possible problems.

\section{Expected hit rates}
\label{sec:hitrate}
The base-line HL-LHC (High Luminosity LHC) scenario foresees an instantaneous luminosity of $5 \times 10^{34}$ cm$^{-2}$s$^{-1}$. Particle fluxes predicted by FLUKA simulation have been convoluted with the RPC sensitivities for a given particle type to obtain the expected hit rate. Two sets of detector sensitivities \cite{uribe, gelmi} have been applied -- for the upgrade iRPC chambers and the ones from the present RPC system. All values are averaged over azimuthal angle $\phi$. A similar approach was followed considering ultimate HL-LHC scenario, where the expected instantaneous luminosity is $7.5 \times 10^{34}$ cm$^{-2}$s$^{-1}$.
\subsection{RPC upgrade region}
\begin{figure}[htbp]
	\centering
	\includegraphics[width=.4\textwidth]{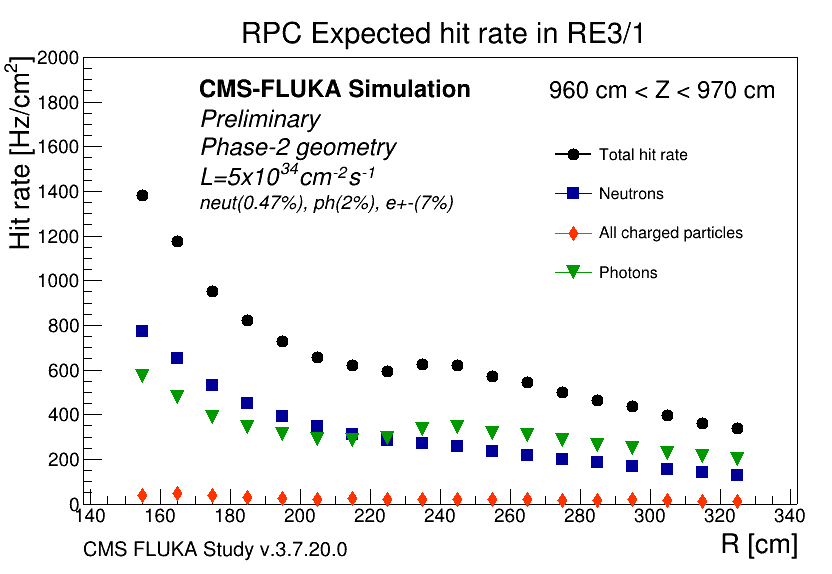}
	\qquad
	\includegraphics[width=.4\textwidth]{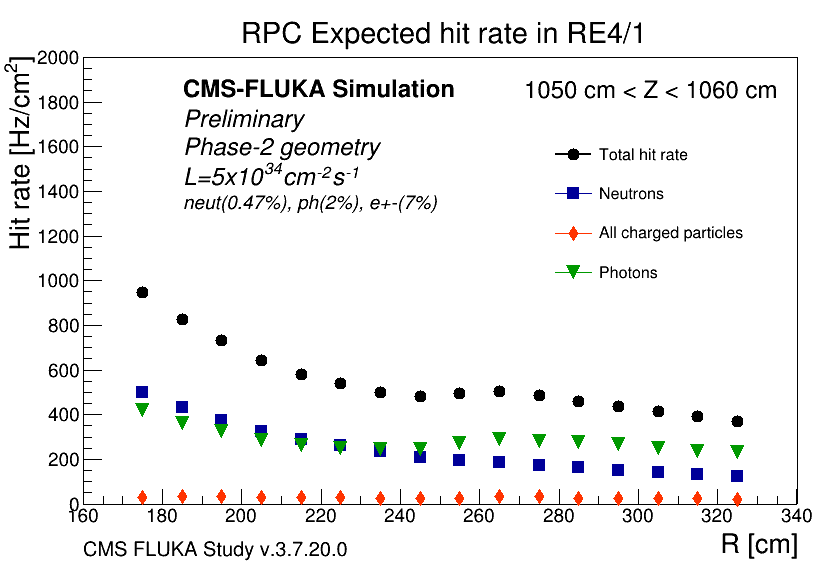}
	\caption{\label{fig:exphitrates} Averaged expected hit rates at an instantaneous luminosity of $5 \times 10^{34}$ cm$^{-2}$s$^{-1}$ are shown in the figure. The plot on the left represents expectation for the RE3/1 iRPC, and on the right, for RE4/1. Different colors depict the contributions from different particle types. Z is along the beam pipe and depicts the distance from the interaction point, and R is a distance from the beam pipe in radial direction.}	
\end{figure}

As can be seen in Fig. \ref{fig:exphitrates} on the left, the highest hit rate is expected at the lower radius of the iRPC chambers of RE3/1 and it is $1400$ Hz/cm$^2$. Averaging over the entire radius range the expected rate is $\sim 660$ Hz/cm$^2$ for RE3/1 and $\sim 500$ Hz/cm$^{2}$ for RE4/1. shown on the right of the same figure. Including a SF (safety factor) of $3$ the values approach $2000$ Hz/cm$^2$ and $1600$ Hz/cm$^2$, respectively. The reported rates are normalized to the chamber area. However, for the choice of detector electronics, the rate per strip is more important. Assuming RE3/1 strips in the range $150$ cm $\leqslant$ R $\leqslant 320$ cm, where R is the distance to the beam pipe, and an average strip pitch of $0.75$ cm (see Ref.\cite{shablo}), the expected rate per strip is $\sim 93$ kHz, applying an SF of $3$ it becomes $\sim 280$ kHz or $0.007$ hits/bx, where $1$ bx $= 25$ ns and corresponds to the detector electronics readout time window. For RE4/1, the expected rate per strip is $0.007$ hits/bx per strip. More details are given in Table \ref{tab:rate_irpc}.

\subsection{Present RPC system}
\begin{figure}[htbp]
	\centering
	\includegraphics[width=.4\textwidth]{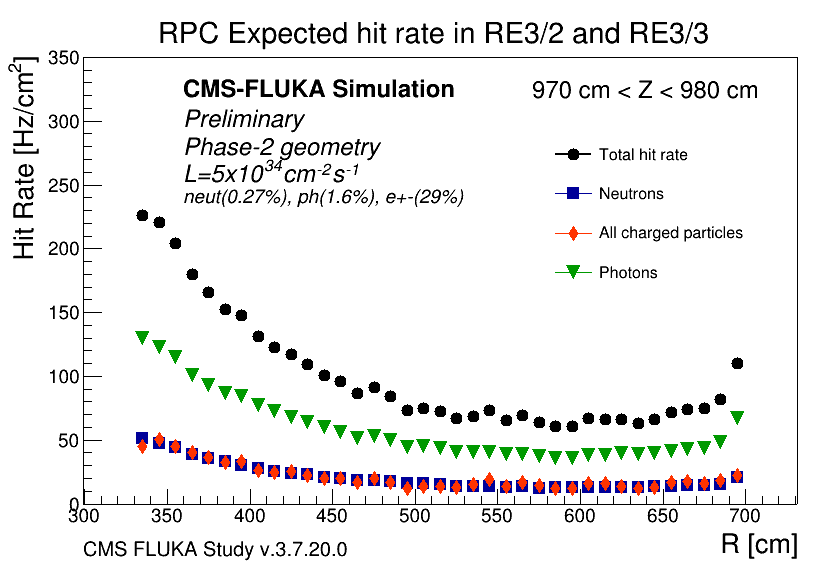}
	\qquad
	\includegraphics[width=.4\textwidth]{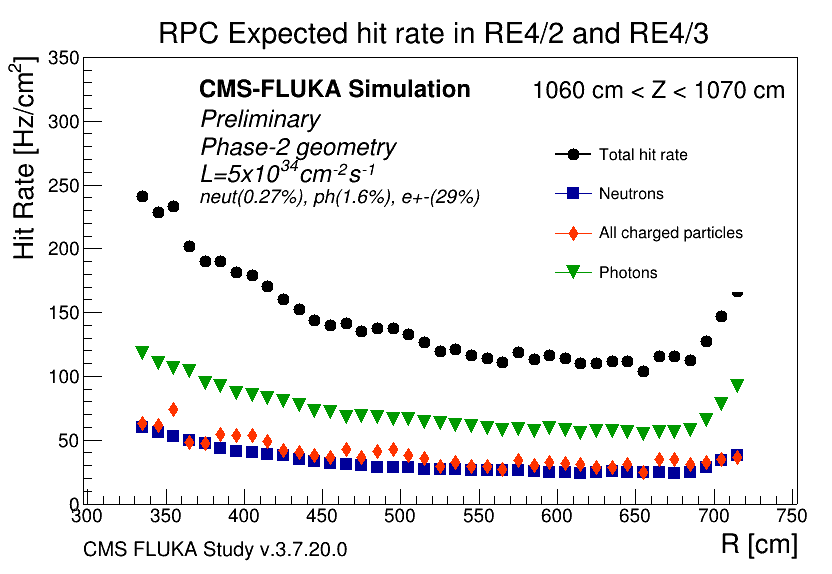}
	\caption{\label{fig:exphitratesPresent} Averaged expected hit rates at an instantaneous luminosity of $5 \times 10^{34}$ cm$^{-2}$s$^{-1}$ are shown in the figure. The plot on the left represents expectation for the RE3/2 and RE3/3 RPC, and on the right for RE4/2 and RE4/3. The second rings RE3/2 and RE4/2 start at $R\sim332$ cm and the third rings RE3/3 and RE4/3 at $R\sim503$ cm. Different colors depict the contributions from different particle types. Z is along the beam pipe and depicts the distance from the interaction point, and R is a distance from the beam pipe in radial direction.}	
\end{figure}

MC predictions at baseline HL-LHC scenario for the third and fourth endcap stations of the present RPC system are shown in Fig. \ref{fig:exphitratesPresent}. While the hit rate increase at the larger radius is caused mainly by the cavern background, the increase at the lower radius is contributed both by the cavern background and the rate of particles from collisions. The highest rate is expected for the outermost stations, which are mainly affected by the neutron background of the cavern. The maximum expected rate is for RE4/2. Averaging along the strip length, and without a safety factor it is $\sim180$ Hz/cm$^2$. This value is comparable with the experimentally derived maximum one ($200$ Hz/cm$^2$) for the RE4/2 top chambers, please see \ref{sec:realdata}. Details for the expected rates and the ultimate HL-LHC scenario are given in Table \ref{tab:rate_irpc}.

\begin{table}[htbp]
	\centering
	\smallskip
	\begin{tabular}{|l|r|r|}
		\hline
	    Average rate $\times$ SF(3) & $L = 5\times10^{34}$cm$^{-2}$s$^{-1}$ & $L = 7.5\times10^{34}$cm$^{-2}$s$^{-1}$\\
		\hline
		RE3/1: in Hz/cm$^2$ & $2000$ & $3000$\\
		RE3/1: per strip in Hz & $280$ & $420$\\
		RE3/1: per strip in hits/bx & $0.007$ & $0.011$\\
        RE4/1: in Hz/cm$^2$ & $1600$ & $2400$\\
        RE4/1: per strip in Hz & $204$ & $300$\\
        RE4/1: per strip in hits/bx & $0.005$ & $0.008$\\        
		RE3/2: in Hz/cm$^2$ & $400$ & $600$\\
		RE4/2: in Hz/cm$^2$ & $540$ & $780$\\
        RE3/3: in Hz/cm$^2$ & $200$ & $300$\\
        RE4/3: in Hz/cm$^2$ & $360$ & $540$\\              
		\hline
	\end{tabular}
	\caption{\label{tab:rate_irpc} Expected rate (scaled by a SF(3)) in the RPC chambers from third (RE3) and forth (RE4) endcap stations at HL-LHC.}
\end{table}

\section{Expected Fluence and Absorbed Dose}
\label{sec:fluensec}
\begin{figure}[htbp]
	\centering
	\includegraphics[width=.325\textwidth]{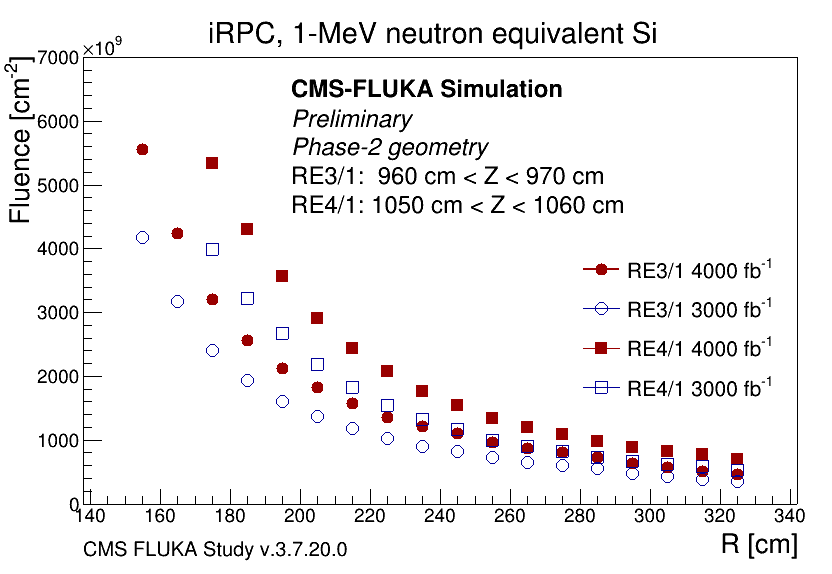}
	\includegraphics[width=.325\textwidth]{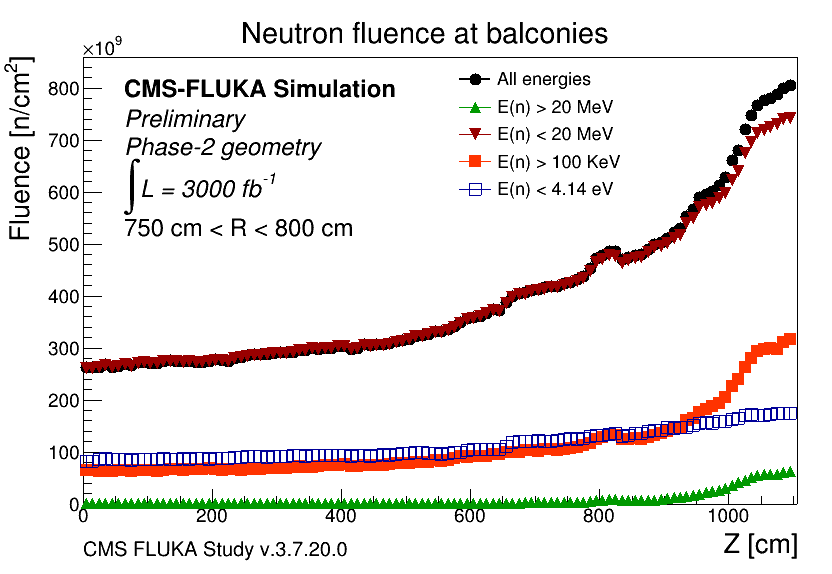}
    \includegraphics[width=.325\textwidth]{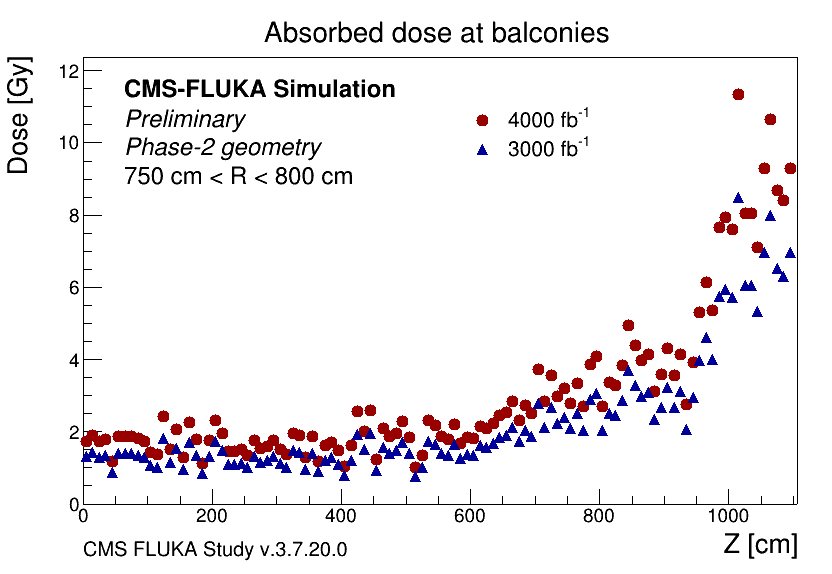}
	\caption{\label{fig:fluencepic} Expected fluence and dose. The plot on the left represents expected fluence (cm$^{-2}$) in terms of 1--MeV neutron equivalent in Si in the iRPC chambers versus the R position, while the plot in the middle represents neutron fluence (n/cm$^{2}$) expected at the balconies in the cavern versus Z position. Absorbed dose after collecting $3000$ fb$^{-1}$ and $4000$ fb$^{-1}$ at the balconies is shown in the plot on the right. SF(3) are not included. Z is along the beam pipe and depicts the distance from the interaction point, and R is a distance from the beam pipe in radial direction.}
\end{figure}

Expected fluence in terms of 1--MeV neutron equivalent in Si for the iRPC chambers is shown in Fig. \ref{fig:fluencepic} on the left. The average ratio between the values obtained with ultimate and base HL-LHC scenarios is $1.33$. Expected fluence at $R=303$ cm for RE3/1 after collecting $3000$ fb$^{-1}$ is $\sim4.3\times10^{11}$ cm$^{-2}$, and at $R=304$ cm for RE4/1 is $\sim6.2\times10^{11}$ cm$^{-2}$, where $R=303$ $(304)$ cm are the expected FEB positions. Part of the RPC electronics is installed far from the apparatus but still inside the cavern - on the balconies of the cavern. Though these places are not reached by particles coming from collisions, the neutron-induced background can affect the electronics and lead to potential damages. The usual place where the RPC racks are installed is between $750$ cm and $800$ cm distance from the beam pipe. The plot in the middle of Fig. \ref{fig:fluencepic} represents the expected neutron fluence at balconies after collecting $3000$ fb$^{-1}$, where the results are provided in ranges of neutron energy. As can be seen, the relative highest contribution comes from neutrons with energy $100$ KeV $ < E < 20$ MeV. Maximum expected fluence in the barrel part ($Z < 600$ cm) is below $350\times10^{9}$ n/cm$^{2}$, while for the endcap ($Z > 600$ cm), the highest expectations are $\sim800\times10^{9}$ n/cm$^{2}$, where  Z is along the beam pipe and depicts the distance from the interaction point.\\
Absorbed dose estimations are biased by the missing details in the geometry description and systematic uncertainties are yet to be assessed. Thus absorbed dose results presented here give only indications about the dose expected values needed for the detector electronic studies. The new results will be reported once the new layers are added to the geometry configuration. The expected absorbed dose after collecting $3000$ $(4000)$ fb$^{-1}$ at the FEB positions is estimated to be $\sim10$ $(13.6)$ Gy for RE3/1 and $\sim18$ $(24)$ Gy for RE4/1. Absorbed dose at balcony regions is shown in Fig. \ref{fig:fluencepic} on the right. The maximum expected absorbed dose in the barrel is less than $2$ Gy, and in the endcap the highest values are below $10$ Gy.

\section{Summary}
The expected radiation background in the CMS RPC system has been studied in terms of expected particle rates,
 absorbed dose and fluence. Two HL-LHC scenarios have been investigated - after collecting $3000$ and $4000$ fb$^{-1}$. 

\acknowledgments
We would like to congratulate and thank our colleagues from the CMS BRIL group, who manage the CMS FLUKA simulation. We would like to thank also all CMS RPC members for their dedicated work, common efforts and shared experience.

\end{document}